\documentstyle[12pt]{article}

\input epsf

\begin{document}

\renewcommand{\thesection}{\arabic{section}.} 
\renewcommand{\theequation}{\thesection \arabic{equation}}
\newcommand{\scs}{\setcounter{equation}{0} \setcounter{section}}
\def\req#1{(\ref{#1})}
\newcommand{\be}{\begin{equation}} \newcommand{\ee}{\end{equation}} 
\newcommand{\ba}{\begin{eqnarray}} \newcommand{\ea}{\end{eqnarray}} 
\newcommand{\la}{\label} \newcommand{\nb}{\normalsize\bf} 
\newcommand{\lb}{\large\bf} \newcommand{\vol}{\hbox{Vol}}
\newcommand{\bb} {\bibitem} \newcommand{\np} {{\it Nucl. Phys. }} 
\newcommand{\pl} {{\it Phys. Lett. }} 
\newcommand{\pr} {{\it Phys. Rev. }} \newcommand{\mpl} {{\it Mod. Phys. Lett. }}
\newcommand{\sg}{{\sqrt g}} \newcommand{\sqhat}{{\sqrt{\hat g}}}
\newcommand{\sqphi}{{\sqrt{\hat g}} e^\phi} 
\newcommand{\sqalpha}{{\sqrt{\hat g}}e^{\alpha\phi}}
\newcommand{\tp}{\cos px\ e^{(p-{\sqrt2})\phi}} \newcommand{\stwo}{{\sqrt2}}
\newcommand{\tr}{\hbox{tr}}

\begin{titlepage}
\renewcommand{\thefootnote}{\fnsymbol{footnote}}

\hfill PUPT--1585

\hfill hep-th/9601003

\vspace{.4truein}
\begin{center}
 {\LARGE D-brane actions}
 \end{center}
\vspace{.7truein}

 \begin{center}

 Christof Schmidhuber\footnote{schmidhu@puhep1.princeton.edu

 \ \ on leave of absence from the Institute for Theoretical Physics, 
University of Bern, Switzerland.}

 \vskip5mm

 {\it Joseph Henry Laboratories}

 {\it Princeton University}

 {\it Princeton, NJ 08544, USA}

 \end{center}

\vspace{.7truein}
\begin{abstract}
Effective world-brane actions for solitons of ten-dimensional type IIA and 
IIB superstring theory
are derived using the formulation of solitons as Dirichlet branes. The 
one-brane actions
are used to recover predictions of $SL(2,Z)$ strong-weak coupling duality.
The two-brane action, which contains a hidden eleventh target space coordinate,
is shown to be
the eleven-dimensional supermembrane action.
It can be thought of as the membrane action of ``M-theory''.

\vfill
\noindent{\it (to appear in Nuclear Physics B)}
\end{abstract}
 \renewcommand{\thefootnote}{\arabic{footnote}}
 \setcounter{footnote}{0}
\end{titlepage}

\section*{1. Introduction}\scs{1}

Over the last year a large amount of evidence has accumulated in support of 
the conjectured
strong-weak coupling duality of superstring theory 
(see, e.g., \cite{lust,sen,hulltown,witt,harstrosen,stro,schwarz}).
According to this conjecture, string theory has a large group of discrete 
symmetries that
exchange elementary strings with solitons and flip the sign of the dilaton, 
thereby transforming
strongly coupled vacua into weakly coupled vacua and vice versa.
A particularly interesting subgroup of the conjectured duality group is the 
$SL(2,Z)$ symmetry group of 
type IIB string theory in ten dimensions. In this case there is an infinite 
set of solitonic strings
which are expected to play the role of elementary strings in the corresponding 
strong-coupling
expansions of the theory \cite{schwarz}.  

Strong-weak coupling duality is a familiar phenomenon in statistical mechanics 
where it is often useful in
extracting information about simple systems like the two-dimensional Ising 
model.
One may hope that duality will likewise be helpful in extracting 
information about nonperturbative aspects of string theory.
A promising step is the
recent formulation of some superstring solitons as ``Dirichlet branes'' 
\cite{polchinski}.
In principle, this should make it possible to directly construct the various 
strong-coupling expansions
of string theory using standard methods of conformal field theory on Riemann 
surfaces with boundaries.

Here this is discussed to lowest order in the string coupling constant.
In particular, the effective world-sheet actions of the $SL(2,Z)$ multiplet of 
solitonic strings
of the type IIB theory are derived from Dirichlet branes and are shown to 
precisely coincide with the
predictions of duality. This requires integrating out an auxiliary gauge field 
that lives on the D-brane.
The method can also be used to discuss other
solitonic $p$-branes that become massless in the strong coupling limit. 
As an example, the two-brane of the ten-dimensional type IIA theory is discussed
which has a hidden eleventh embedding dimension and can be thought of as the 
membrane of
the compactified so-called ``M-theory''.\footnote{M-theory 
has been suggested as a kind of master theory which contains 
eleven-dimensional supermembranes
and from which many of the string dualities can be derived 
\cite{schwarz,horava}.}
The world-brane action is demonstrated to be the standard eleven-dimensional 
supermembrane action \cite{tow}.\footnote{The reverse,
namely deriving the two-brane action from the supermembrane action, has 
recently been suggested in \cite{townsend}. The identification of the
type IIA two-brane action with the eleven-dimensional supermembrane
action has first been suggested in \cite{duff}.}

How to use D-branes to compute effective world-brane actions for general 
$p$-brane solitons in flat
ten-dimensional Minkowski space is briefly explained in section 2. This direct 
extension
of the calculation of the one-loop effective action of the type I string 
\cite{clny} relies on previous work 
\cite{witten,li,calkle} and reproduces and generalizes 
results obtained with different methods in \cite{leigh,chs}.
We focus on the bosonic part of the action (the fermionic part is determined 
by supersymmetry).
The one-branes (soliton strings) are discussed in section 3 and the 
two-branes are discussed in section 4. 
Section 5 contains conclusions and summarizes open ends.
We plan to give a more exhaustive discussion of the actions of solitonic 
$p$-branes in the future. 

While this work was being completed,
discussions with partial overlap appeared \cite{douglas,townsend}.

\section*{2. World-brane actions as string loop corrections}\scs{2}

In this section, some facts about D-branes, type II strings and boundary 
states are briefly reviewed
and assembled to yield world-brane actions of $p$-brane solitons. 

To study superstring solitons in conformal field theory, one adds
a boundary $\partial\Sigma$ to the world-sheet $\Sigma$ of a type II string. 
One assumes that $p+1$ coordinates $x^\alpha$ with $\alpha\in\{0,...,p\}$ obey 
Neumann boundary 
conditions, while $9-p$ coordinates $x^i$ with $i\in\{p+1,...,9\}$ obey 
Dirichlet boundary conditions. 
Thus the boundary is restricted to lie inside a $(p+1)$-dimensional submanifold 
of the target space, the ``Dirichlet brane'' or ``D-brane'' \cite{dlp}. 
Conformal invariance of the boundary theory requires the D-brane to obey 
certain equations of motion \cite{leigh}.

The massless field of the boundary theory is
a $U(1)$ super Yang-Mills target space field.
The bosonic parts of the corresponding vertex operators are
\ba\oint_{\partial\Sigma} \{\ A_{i}(x)\ \partial_\sigma x^i \}
\ \ \ \ \hbox{and}\ \ \ \ \oint_{\partial\Sigma} \{\ A_{\alpha}(x)\ 
\partial_\tau x^\alpha
  \}\ \la{one}\ea
where $\partial_\sigma$ and $\partial_\tau$ denote the derivatives
normal and tangential to the world-sheet boundary, respectively.
As explained in \cite{dlp,leigh}, the
first type of operator corresponds to transverse excitations of the D-brane.
Those will be ignored in the following, since they can always be absorbed in 
position and shape of the D-brane.

The second type of operator corresponds to an internal gauge field that lives 
on the D-brane.
It can be regarded as an auxiliary field needed to insure gauge invariance of 
the Neveu-Schwarz
antisymmetric tensor field $B_{\alpha\beta}$, pulled back onto the D-brane:
under a gauge transformation $B_{\alpha\beta}\rightarrow  B_{\alpha\beta}
+\partial_{[\alpha} \Lambda_{\beta]}$, the change of the world-sheet action is
$$\ \int\epsilon^{\gamma\delta}\ \partial_\gamma x^\alpha\partial_\delta 
x^\beta\ \partial_{[\alpha}\Lambda_{\beta]}
   \ =\ \ \oint \Lambda_\alpha\ \partial_\tau x^\alpha\ ,$$
which amounts to a transformation of the boundary gauge field,
$$ A_\alpha\rightarrow A_\alpha+\Lambda_\alpha\ \ \ \ \ \Longrightarrow\ \ \ \ \
    F_{\alpha\beta}\rightarrow F_{\alpha\beta}
+\partial_{[\alpha}\Lambda_{\beta]}\ .$$ 
The modified field strength of the boundary gauge field $A_\mu$,
$${\cal F}_{\alpha\beta}\ =\ F_{\alpha\beta}\ -\ B_{\alpha\beta}\ ,$$
is then the gauge invariant quantity that will appear in the effective 
world-brane action.
The path integral over $A$ can be thought of as an integral over the gauge 
group of $B$.

The D-brane describes a soliton solution of string theory in the sense that 
it is a source for the closed string fields,
which obey the standard closed string equations of motion outside the D-brane.
As explained in \cite{polchinski}, adding
the boundary to the world-sheet breaks half of the supersymmetry of the 
type IIB theory, which implies that the D-brane
corresponds to a BPS-saturated soliton state. Its classical tension formula 
$T\sim{1\over\kappa}$
\cite{dlp} can then be trusted even for strong coupling. This implies that
the mass per unit volume of the brane goes to zero in the strong coupling limit.

Note that while the D-brane is a classical object in 
target space if only one Dirichlet boundary is considered,
the D-brane becomes a quantum mechanical object if 
we sum over arbitrary numbers of world-sheet boundaries.\footnote{For a 
discussion of the divergences associated with multiple
Dirichlet boundaries, see \cite{green}.}
This is the open-string version of the familiar fact that 
summing over world-sheet topologies turns the target space
fields into quantum operators. In this case, the target space 
fields are the fields (\ref{one}): the
(transverse) coordinates of the D-brane and the internal gauge field.

What is meant in the following by the ``effective world-brane action of the 
D-brane'' is simply the one-loop correction
to the effective action of type II string theory that 
arises from adding a Dirichlet boundary
to the string world-sheet. This loop correction is an integral over the 
$p$-brane that arises
as a variant of the well-known Fischler-Susskind effect \cite{fs}:
in the limit where the Dirichlet boundary becomes very small, the 
world-sheet can be conformally mapped
onto a disk connected to a two-sphere by a long thin tube whose length $L$
is a modulus that must be integrated over (fig. 1). 
This integral is logarithmically divergent due to massless closed 
string states $|\Phi_i>$ propagating through the tube.
Those states can be thought of as the components of a ``boundary state'' 
$$|B>\ =\ c^i\ |\Phi_i>$$
that is created by the boundary.
In the standard fashion, the divergences imply contributions 
proportional to $c^i$ to the beta functions $\beta^i$ of the world-sheet 
theory.\footnote{For massless 
modes with weight $h_i=2$, the integral $\int dL\ e^{-L(h_i-2)}$ over the 
modulus $L$ is $\sim\log a$, where $a$ is a short-distance
cutoff on the world-sheet. This dependence on 
the cutoff scale contributes to the beta functions.}
Since the boundary is confined to the D-brane,
the corresponding divergences and contributions to the beta functions 
are $(p+1)$-dimensional integrals over 
the D-brane,
rather than integrals over all of target space. 

{}\epsffile[10 260 0 0]{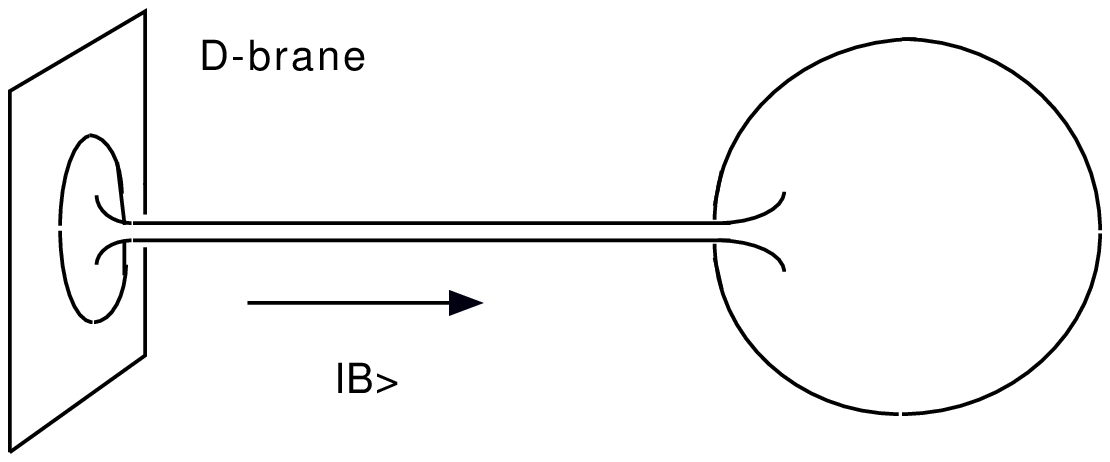}
\vskip5cm

\begin{center} Fig. 1 \end{center}

Without loop corrections, the string equations of motion 
are simply $\beta^i=0$, where $\beta^i\sim{\delta S\over\delta\Phi_i}$
are the genus-zero beta functions. The loop corrected equations of motion are
$$ \beta^i\ +\ \beta^i_{loop}\ =\  \beta^i\ +\ c^i\ \delta(\Sigma_p)\ =\ 0$$
where $\delta(\Sigma_p)$ denotes the restriction to the $p$-brane $\Sigma_p$.
These loop-corrected equations of motion are solved by weighing the 
D-brane with the exponential of a loop-corrected
action
\ba\exp\{\ -\int_\Sigma d^{p+1}x\ {\cal L}_{loop}\ \}\ \ \ \ \ \hbox{with}\ \ \ \ \ {\delta {\cal L}_{loop}\over\delta\Phi_i}\ =\ c^i\ .\la{bob}\ea

The components $|\Phi_i>$ of the boundary state come from 
both the Neveu-Schwarz--Neveu-Schwarz (NS-NS) and the Ramond-Ramond (RR)
sectors.
In the type II string, the massless fields of the
NS-NS sector are the metric $g_{\mu\nu}$, the two-form $B_{\mu\nu}$ and 
the dilaton $\phi$.
The massless fields of the RR sector are $q$-forms $A_q$ with 
duals $*A_q\sim A_{8-q}$, where
$q\in\{0,2,4,6,8,10\}$ for the type IIB string and $q\in\{1,3,5,7,9\}$ for 
the type IIA string.
$A_9$ and $A_{10}$ have no propagating degrees of freedom.

Let us first recall a few relevant facts about type II strings.
The classical low-energy effective string action 
is\footnote{No action can be written for $F_5\neq0$.} (setting $\alpha'=1$):
\ba S &=&  \int d^{10}x\sg e^{-2\phi}\{R+4(\nabla\phi)^2-{1\over12} H^2 \}\\ 
&&\ -\ \sum_k\ 
\int d^{10}x\ \sg\ {1\over2\ k!}F_k^2\ +\ \hbox{cubic terms}\ \ea
where $H=d\wedge B$ is the field strength of $B$ and 
the field strengths $F_k$ are defined as
$$ F_1\ =\ d\wedge A_0\ ,\ \ \ F_2\ =\ d\wedge A_1\ ,\ $$
etc.\footnote{For $k>2$, the definition of $F_k$ contains 
$H$-dependent terms that will not be important here.}
These definitions will have to be modified below, when loop corrections 
are taken into account.
As they are now, they lead to the Bianchi identities
$$ d\wedge F_1\ =\ 0\ ,\ \ \ d\wedge F_2\ =\ 0\ ,\ $$
etc.
The potentials are defined here such that the gauge transformations 
are ``natural'':
$$ \delta A_{q+1}\ =\ d\wedge\Lambda_q$$
with $q$-forms $\Lambda_q$. Redefining $A_{q+1}\rightarrow e^{\phi}A_{q+1}$ 
would add
the familiar factor $e^{-2\phi}$ to the RR part of the 
effective action but make the transformation law
$\phi$-dependent \cite{witt}.

The beta functions of the two-dimensional sigma model are the standard 
beta functions for the NS-NS fields,
$$ \beta^g_{\mu\nu}\ =\ R_{\mu\nu}+2\nabla_\mu\nabla_\nu\phi+...\ ,\ $$
etc., and
$$ \beta^{A_q} \ \sim\ d\cdot F_{q+1}+...$$
for the RR fields, where $d\cdot F$ denotes $\nabla^\mu F_{\mu\mu'...}$.

Let us now symbolically write the
boundary state that is created by a Dirichlet boundary of a type II 
string world-sheet as
\ba   |B> &\sim& c^i|\Phi_i>\ =\
   c^h_{\mu\nu}|h^{\mu\nu}> +c^B_{\mu\nu}|B^{\mu\nu}> +c^\phi|\phi>+
 \sum_q c^{A_q} |A_q>\ .\ea
Here the notation is somewhat sloppy but sufficient for our 
purposes: $|A_q>$ denotes the state that
is created by the corresponding RR vertex operator (which in fact 
depends only on the field strength of $A_q$); see \cite{cai,clny,li} for 
details.
$|B>$ in the presence of the boundary gauge field $A_{\alpha}$ has been 
discussed in \cite{li,calkle}. The analysis is valid
for slowly varying field strengths ${\cal F}$ as defined above\footnote{There
are corrections to the following formulas of order $(\nabla^2{\cal F})^2$.} 
and follows 
that for the type I string \cite{clny}. 

In the NS-NS sector,
the graviton, $B$-field and dilaton components 
of the boundary state correspond to the
symmetric part, antisymmetric part and the trace of a $(p+1)$ by $(p+1)$ matrix
that ``rotates'' the left- and right-moving components of 
the ${\cal F}=0$ boundary state relative to each other
($G$ denotes the induced metric on the D-brane; one expands around $G=\eta$):
\ba |B>_{NS} &\sim& e^{-\phi}\ {\sqrt{-\det (G+{\cal F})}}\\
&\times& \{\ ({G-{\cal F}\over G+{\cal F}})^{\alpha\beta}
  \ (\ |h_{\alpha\beta}>\ +\ |B_{\alpha\beta}>\ )\ -\ 2|\phi>\ \}\ ,\la{thr}\ea
The action in (\ref{bob}) then becomes
\ba S\ =\  \int_\Sigma d^{p+1}x\ e^{-\phi}\ {\sqrt{-\det(G+{\cal F})}} \ .
 \la{jan}\ea
This is the Dirac-Born-Infeld action first derived for bosonic D-branes 
in \cite{leigh} (and for open strings in \cite{fratsey});
see, e.g., \cite{bachas,calkle} for discussions. 

The massless RR part of the boundary state is a linear 
combination of the states that are obtained
by acting with the anticommuting fermionic 
zero modes $\theta_0^\mu = \psi_0^\mu+\tilde\psi_0^\mu$
of the RR sector on the RR ground state $|0>$ 
(we omit the ghosts for simplicity of notation
and now also convert to the formalism where $A_q$ is rescaled by the factor 
$e^\phi$)\footnote{One could also use a dual notation,
where an $n$-form 
corresponds to $\theta_0^{\mu_1}...  \theta_0^{\mu_{n}}\ |0>$.}:
\ba \theta_0^{\mu_1}...  \theta_0^{\mu_{n}}\ |0>\ 
 &\sim&\ {1\over(10-n)!}\ \epsilon^{\mu_1...\mu_{10}}
     |(A_{10-n})_{\mu_{n+1}...\mu_{10}}>\ .\la{two}\ea
Here, the state $|A_{8-q}>\equiv |*A_{q}>$ is the dual of the state $|A_q>$.
The dual states in $|B>$ (with $q\in\{4,5,6,7,8\}$) 
modify the Bianchi identities to
$$d\wedge F_{9-q}\ \sim\ c^{q}\ \delta(\Sigma_{p})\ $$
(plus $H$-dependent terms).
This is analogous to the fact that in Maxwell theory the Bianchi 
identity $d\wedge F=0$ is not globally valid in the presence
of a magnetic monopole: $d\wedge F$ has delta-function support at the 
location of the monopole.
In the case at hand, the $(8-p)$-brane is a source of $F_p$ magnetic charge.

For ${\cal F}=0$, the boundary state is \cite{polchinski}
\ba|B>_{{\cal F}=0}\ \sim\ \pm\ \theta_0^{p+1}...\theta_0^9\ |0>\ .\la{fiv}\ea
This is just the statement that the $p$-brane carries the corresponding 
$p$-form charge (the 
sign ambiguity corresponds to D-branes versus anti-D-branes \cite{li}).
Turning on ${\cal F}$ rotates the RR boundary state
simultaneously with its NS-NS counterpart in (\ref{thr}) because of 
supersymmetry.
The spinor-spinor analog of the ``rotation operator'' in (\ref{thr}) acts 
as \cite{clny}:
\ba |B>_{RR} &\sim& \tr\ e^{- {1\over2}\theta_0^\alpha\theta_0^\beta\ 
{\cal F}_{\alpha\beta}}|B>_{{\cal F}=0}\\
             &\sim& \tr\{\ 1\ -\ {1\over2}\theta_0^\alpha\theta_0^\beta\ 
{\cal F}_{\alpha\beta}
   +\ {1\over8}\ \theta_0^\alpha\theta_0^\beta\theta_0^\gamma\theta_0^\delta\
       {\cal F}_{\alpha\beta}{\cal F}_{\gamma\delta}
             +... \ \}|B>_{{\cal F}=0}\ .\la{fou}\ea
This expansion in $\theta$ terminates when the number 
of $\theta_0$'s acting on $|B>_{{\cal F}=0}$ 
reaches $p+1$.
 From (\ref{two},\ref{fiv},\ref{fou}) the coefficints $c_i$ 
of the boundary state can now be
simply read off.\footnote{To compare with the type 
I string, type I string loop corrections are obtained by setting $p=9$, changing
the gauge group to $SO(32)$ and dropping odd powers of ${\cal F}$ which is 
equal to $F$ since there is no NS-NS $B$-field.
The ${\cal F}$-independent part of the boundary state is then cancelled by an 
opposite contribution from
the crosscap (in our case there is no crosscap since type II strings are 
oriented).}

The states $|A_q>$ with $q\in\{0,1,2,3,4\}$ modify the equations of motion, 
yielding the world-brane action (see (\ref{bob}))
$$\int_\Sigma d^{p+1}x\ c^i\ \Phi_i\ .\ $$
Let us illustrate this at the examples $p=1$ and $p=2$.
For the one-brane of the type IIB theory one has $\alpha,\beta\in\{0,1\}$. 
The RR part of the boundary state is
$$(1-{1\over2}\theta_0^\alpha\theta_0^\beta{\cal F}_{\alpha\beta})\ 
\theta_0^2...\theta_0^{10}\ |0>\
     =\ {1\over2}\epsilon^{\alpha\beta}(\ |(A_2)_{\alpha\beta}>\ -\ 
|A_0>{\cal F}_{\alpha\beta}\ )\ .$$
This yields the effective action (with $A^{(k)}\equiv A_k$)
\ba {1\over2}\int d^2x\ \epsilon^{\alpha\beta}(A^{(2)}_{\alpha\beta}
-A^{(0)}\ {\cal F}_{\alpha\beta})\ .\la{feb}\ea
For the two-brane of the type IIA theory, one has $\alpha,\beta\in\{0,1,2\}$. 
The RR part of the boundary state is
$$(1-{1\over2}\theta_0^\alpha\theta_0^\beta{\cal F}_{\alpha\beta})\ 
\theta_0^3...\theta_0^{10}\ |0>\
     =\ {1\over2}\epsilon^{\alpha\beta\gamma}(\ 
 {1\over3}|A^{(3)}_{\alpha\beta\gamma}>\ 
-\ {\cal F}_{\alpha\beta}|A^{(1)}_\gamma>\ )\ .
$$
This yields the effective action
\ba{1\over2}\int d^3x\ \epsilon^{\alpha\beta\gamma} ({1\over3}
A^{(3)}_{\alpha\beta\gamma}-A^{(1)}_\alpha{\cal F}_{\beta\gamma})\ .\la{tue}\ea
These cases will be discussed in detail in the next sections.

\section*{3. One-branes}\scs{3}

For the solitonic strings of the type IIB theory, 
(\ref{jan}) and (\ref{feb}) combined yield the effective world-sheet action
(in Minkowski space; we adopt the more common 
notation $B\equiv B^{(1)}, A^{(0)}\equiv\chi, A^{(2)}\equiv B^{(2)}$):
\ba S\ =\  \int d^2x\ n\ \{\ e^{-\phi}\ {\sqrt{-\det(G+{\cal F})}} \
 +\ {1\over2}\ \epsilon^{\alpha\beta} B^{(2)}_{\alpha\beta}\
 -\ {1\over2}\chi\ \epsilon^{\alpha\beta}{\cal F}_{\alpha\beta}\ \}\ .
 \la{mar}\ea
Here we have included a factor $n$ to describe the bound state of 
$n$ soliton strings.\footnote{This corresponds
to replacing the two-dimensional $U(1)$ gauge theory with a $U(n)$ gauge 
theory that has a mass gap \cite{witten}; because of the gap, the trace
over the group indices effectively reduces to the factor $n$.}
By comparison, the standard Nambu-Goto action of 
an elementary string with tension $\tilde T$ in an NS-NS two-form background
 $\tilde B^{(1)}$ is
\ba S\ =\  \int d^2x\ \{\ \tilde T\ {\sqrt{-\det\ G}} \
 +\ {1\over2}\ \epsilon^{\alpha\beta} \tilde B^{(1)}_{\alpha\beta}\ \}\ .
 \la{apr}\ea
If we want to interpret the soliton string as an 
elementary string, we must integrate out the auxiliary gauge field $A$
 that enters through ${\cal F}$ in (\ref{mar}). If duality 
is correct, then this should yield an action of the form (\ref{apr}),
with modified background fields such as ``seen'' by the soliton string.

$A$ will be integrated out here in a way that can later be generalized 
to higher $p$-branes.
First, the Born-Infeld part of (\ref{mar}) is rewritten 
with the help of an auxiliary scalar field $t$:
\ba &&\exp\{\ -\int d^2 x\ n\ e^{-\phi}\ {\sqrt{-\det(G+{\cal F})}} \ \}
\la{jun}\\
    && = \int {\cal D} t\ \exp\{\ -\int d^2x\ [ \ e^{-\phi}\ 
 {\sqrt{(-\det\ G)(n^2+t^2\ e^{2\phi})}}\ +\ {1\over2}t\ 
\epsilon^{\alpha\beta}{\cal F}_{\alpha\beta}\ ]\ \}\ .\la{jul}\ea
Indeed, since $t$ does not propagate, we replace it in the action  
(\ref{jul}) by its saddle point value\footnote{since there are no
universal, logarithmically divergent loop contributions to the effective
action.}
which is determined by
\ba
   e^{\phi}\ {\sqrt{-\det\ G}}\ {1\over{\sqrt{n^2+t^2\ e^{2\phi}}}}\ t\ 
=\ -{1\over2}\ \epsilon^{\alpha\beta}{\cal F}_{\alpha\beta}\ .\la{may}\ea
After some straightforward algebra this yields (\ref{jun}).\footnote{It 
is helpful to use a diffeomorphism to set $G=({}^{-g}\ {}_g)$
and to define ${\cal F}=({}_{-f}\ {}^f)$ such that $-\det(G+{\cal F})=g^2-f^2$,
 and to once square (\ref{may})
and once multiply (\ref{may}) by $t$.}

The complete action (\ref{mar})+(\ref{jul}) is then
$$ \int d^2x\ \{\ e^{-\phi}\ {\sqrt{(-\det\ G)(n^2+t^2\ e^{2\phi})}}\ 
+\ {n\over2}\ \epsilon^{\alpha\beta} B^{(2)}_{\alpha\beta}\ 
-\ {1\over2}(n\chi-t)\ 
\epsilon^{\alpha\beta}(F_{\alpha\beta}-B^{(1)}_{\alpha\beta})\ \}\ .$$
The gauge field $A_\alpha$ in $F_{\alpha\beta}$ now 
acts as a Lagrange multiplier that enforces the constraint
$$n\chi\ -\ t\ =\ \ \hbox{constant}\ \ \equiv\ m\ .$$
If the coordinate $x_1$ is compactified on a circle, $m$ is an integer 
\cite{witten}.\footnote{because of the sum over large gauge transformations;
the integral over the constant mode of $t$ then becomes a sum over $m$.}
Integrating out $A_\alpha$, substituting for $t$ and writing $\kappa=e^\phi$ 
now yields the world-sheet action
\ba \int d^2x\ \{\ {\sqrt{{n^2\over\kappa^2}+(m-n\chi)^2}}\ {\sqrt{-\det\ G}}\ 
+\ {1\over2}\ \epsilon^{\alpha\beta}(\ n\ B^{(2)}_{\alpha\beta}
    +\ m\ B^{(1)}_{\alpha\beta}\ )\ \}\ .\la{sep}\ea
This is indeed of the form (\ref{apr}) with the dual tension and $B$-field 
given by
\ba \tilde T &=& {\sqrt{{n^2\over\kappa^2}+(n\chi-m)^2}}\la{aug}\\
 \tilde B^{(1)} &=& mB^{(1)}\ +\ nB^{(2)}\ .\ea
These are precisely the predictions of $SL(2,Z)$ duality \cite{schwarz}: 
there is an infinite set of soliton strings
with charges $(m,n)$ with respect to the NS-NS and RR two-forms $B^{(1)}$ and 
$B^{(2)}$ ($(1,0)$ is the elementary
string) and with tensions (\ref{aug}). Each 
of these soliton strings gives rise to a string perturbation
expansion. Indeed, as mentioned in section 2, allowing 
for an arbitrary number of Dirichlet boundaries on the
world-sheet will turn the positions $A^i$ of the soliton string into 
quantum variables. So the path integral
over Riemann surfaces with (Dirichlet) boundaries implicitly contains 
path integrals over soliton strings. 

The effective world-sheet actions (\ref{sep}) have been derived only to lowest 
order in $\kappa$ and $\alpha'$. A priori, it is not clear that that they can be taken
seriously at higher orders. Remarkably, if duality is correct then
those actions must in fact be exact. In particular, the dual dilaton $\tilde\phi$ (whose expectation value is the dual string coupling 
constant) determines how
soliton strings
of higher genus are weighted. $\tilde\phi$ can be obtained by absorbing
$\tilde T$ in (\ref{apr}) in a dual metric $\tilde G$ and by noting that the
dual Einstein metric $\tilde G^E_{\mu\nu} =\ \tilde 
G_{\mu\nu}e^{-{\tilde\phi\over2}}$
must remain unchanged, since it is defined to 
be \ $\sim\eta_{\mu\nu}+h_{\mu\nu}$:
$$ G^E_{\mu\nu} =\ G_{\mu\nu}e^{-{\phi\over2}}\ 
=\ \tilde G_{\mu\nu}e^{-{\tilde\phi\over2}}\ =\
\tilde G^E_{\mu\nu}\ .$$
It then follows from $\tilde G\ =\ \tilde T\ G$ that the dilaton transforms as
$$ e^{-\tilde\phi}\ =\ {1\over\tilde T^2}e^{-\phi}\ 
=\ e^{-\phi}{1\over n^2e^{-2\phi}+(m-n\chi)^2}\ 
=\ e^{-\phi}{1\over n^2 \vert \lambda-{m\over n}\vert^2}\ $$
where the complex field
$$\lambda = \chi+ie^{-\phi}$$
has been defined. This again agrees with the duality prediction \cite{schwarz} 
which can be seen to be in this case 
$$\hbox{Im}\ \tilde \lambda\ =\ \hbox{Im}\ {-1\over n^2(\lambda-{m\over n})}\ 
.$$
In particular, for $m=0,\chi=0$ the string coupling constant is inverted:
$$ \tilde\kappa\ =\ {n^2\over\kappa}\ .$$

Duality makes predictions not only for the dual NS-NS backgrounds 
$\tilde B^{(1)},\tilde G,\tilde\phi$ but also for the dual RR backgrounds
$\tilde B^{(2)},\tilde\chi$. E.g., for $B^{(2)}=0,m=0$:
$$  \tilde B^{(2)}_{\mu\nu} \ =\  -{1\over n}B^{(1)}_{\mu\nu}\ .$$
It would be very interesting (but is left for future work) 
to derive the dual RR backgrounds directly from D-branes. 
Perhaps this can teach us a simple way
to represent RR backgrounds on the string world-sheet.

\section*{4. Two-branes}\scs{4}

As another example, let us
consider the two-brane of the type IIA theory. What 
makes this two-brane particularly interesting is that it can be thought of as
the membrane of M-theory (the ``M-brane'' \cite{townsend}).\footnote{M-theory 
is the hypothetical strong-coupling limit
of type IIA string theory whose low-energy limit is eleven-dimensional 
supergravity \cite{witt}. 
Among other objects, M-theory must contain supermembranes.}
After compactification on a circle from eleven to ten dimensions,
$M$-branes that are not wrapped around the circle are supposed to 
become the RR-charged two-branes
of type IIA string theory. This will now be used to derive the 
M-brane action from Dirichlet two-branes.\footnote{The reverse 
has already been suggested in \cite{townsend}: assuming that the M-brane 
action is the standard supermembrane action
one can derive the D-brane action. This has been demonstrated 
in \cite{townsend} for part of the action in the
limit where the Born-Infeld action can be approximated by a Yang-Mills action.}

The combined effective type IIA two-brane action (\ref{jan})+(\ref{tue}) is
\ba S\ =\  \int d^3x\ \{\ e^{-\phi}\ {\sqrt{-\det(G+{\cal F})}} \
 +\ {1\over6}\ \epsilon^{\alpha\beta\gamma} A^{(3)}_{\alpha\beta\gamma}\
 -\ {1\over2}\ \epsilon^{\alpha\beta\gamma}C_\alpha{\cal F}_{\beta\gamma}\ \}\ 
 .\la{fri}\ea
Here, $\alpha,\beta\in\{0,1,2\}$, and $A^{(1)}_\alpha$ has 
been renamed $C_\alpha$ to avoid confusion with the
boundary gauge field $A_\alpha$.
As before, $A_\alpha$ will be regarded as an auxiliary gauge field that 
will be integrated over. 
This will be seen to produce an eleventh embedding
dimension and a membrane action that posesses 
eleven-dimensional general covariance, namely the standard eleven-dimensional
supermembrane action \cite{tow}, as first suggested in \cite{duff}.

In analogy with the last section, let us rewrite the Born-Infeld action 
with the help of an auxiliary vector field
$t_\alpha$ (compare with \cite{townsend}):
\ba && \exp\{-\int d^3x\ e^{-\phi}\ {\sqrt{-\det(G+{\cal F})}}\ \}\la{oct}\\
    && =\ \int {\cal D}t_\alpha\ \exp\{ -\int d^3x\ [\ e^{-\phi}\ 
 {\sqrt{-\det(G_{\alpha\beta}+e^{2\phi}\ t_\alpha t_\beta)}}\ +\ {1\over2}
\epsilon^{\alpha\beta\gamma} t_\alpha{\cal F}_{\beta\gamma}\ ]\ \}\ .\la{nov}\ea
To see that (\ref{oct}) and (\ref{nov}) are equivalent, 
one first notes that\footnote{It 
is useful to diagonalize $G$ by a diffeomorphism; note that $G_{00}<0$.}
\ba
-\det(G+{\cal F}) &=& (-\det G)\ (1\ +\ {1\over2}
  G^{\alpha\gamma}G^{\beta\delta}{\cal F}_{\alpha\beta}
 {\cal F}_{\gamma\delta})\ \la{sat}\\
-\det(G_{\alpha\beta}+e^{2\phi}\ t_\alpha t_\beta) &=& (-\det G)\ (1\ 
+\ e^{2\phi}\ G^{\alpha\beta}\ t_\alpha t_\beta)\ .\la{sun}\ea
Here and below, the metric $G$ is used to raise and 
lower indices. Again, we replace the nonpropagating field $t_\alpha$ 
in (\ref{nov}) by its saddle point value. The saddle point equation is
\ba  e^{\phi}\ {\sqrt{-\det\ G}}\ {1\over {\sqrt{1+e^{2\phi}\ 
t^\alpha t_\alpha}}}\ t^\alpha\ =\ -{1\over2}\ 
\epsilon^{\alpha\beta\gamma}{\cal F}_{\beta\gamma}\ .\la{dec}\ea
After once squaring (\ref{dec}) and once multiplying (\ref{dec}) 
with $t_\alpha$, and by using (\ref{sat},\ref{sun}), 
it is indeed straightforward to see that the saddle point
value of the action in (\ref{nov}) is the action in (\ref{oct}). 

The complete action (\ref{fri})+(\ref{nov}) is 
\ba \int d^3x\ \{\ e^{-\phi}\ {\sqrt{-\det(G_{\alpha\beta}+e^{2\phi}\ 
t_\alpha t_\beta)}}\ +\ {1\over6}\epsilon^{\alpha\beta\gamma} 
   A^{(3)}_{\alpha\beta\gamma}\ -\ {1\over2}\epsilon^{\alpha\beta\gamma}
(C_\alpha-t_\alpha)(F_{\beta\gamma}-B_{\beta\gamma})\ \}\ .\la{mon}\ea
As in the case of one-branes, the boundary 
gauge field $A$ in $F_{\beta\gamma}=\partial_{[\beta}A_{\gamma]}$ 
now acts as a Lagrange multiplier.
Integrating it out enforces the constraint
$$\epsilon^{\alpha\beta\gamma}\ \partial_\beta\ (C_\gamma\ -\ t_\gamma)\ 
=\ 0\ .$$
This allows one to write 
$$C_\alpha\ -\ t_\alpha\ \equiv\ \partial_\alpha\ y\ $$
with a scalar field $y$. Substituting for $t_\alpha$ in (\ref{mon}) 
and pulling the factor $e^{-\phi}$ under the square root
finally yields the world-brane action\footnote{not writing out determinants
from the change of variables}
\ba S\ =\ \int\ d^3x\ \{\ {\sqrt{-\det\ \hat G}} \ +\ {1\over6}\ 
\epsilon^{\alpha\beta\gamma}\hat A^{(3)}_{\alpha\beta\gamma}\ \}\la{wed}\ea
with 
\ba
\hat G_{\alpha\beta} &=& G_{\alpha\beta}\ e^{-{2\over3}\phi}\ +\ e^{{4\over3}
\phi}\ (\partial_\alpha y-C_\alpha)\ (\partial_\beta y-C_\beta)\ ,\\
\hat A^{(3)}_{\alpha\beta\gamma} &=& A^{(3)}_{\alpha\beta\gamma}\ 
+\ 3\ \partial_\alpha y\ B_{\beta\gamma}\ .\ea
The induced background fields can be written as
\ba  G_{\alpha\beta} &=& \partial_\alpha x^i\partial_\beta x^j\ G_{ij}\\  
 B_{\alpha\beta} &=& \partial_\alpha x^i\partial_\beta x^j\ B_{ij}\\ 
  A^{(3)}_{\alpha\beta\gamma} &=& \partial_\alpha x^i\partial_\beta x^j 
\partial_\gamma x^k\ A^{(3)}_{ijk}\\ 
  C_{\alpha} &=& \partial_\alpha x^i\ C_i\ .\ea
One sees that $y$ acts as an additional target 
space coordinate and that the type IIA target space fields
combine into the metric $\hat G$ and the three-form $\hat A^{(3)}$ 
of eleven-dimensional supergravity as usual:
\ba {\hat G}_{\mu\nu}\ dx^\mu dx^\nu &=& e^{-{2\over3}\phi}\ G_{ij}\ dx^idx^j\
  +\ e^{{4\over3}\phi}\ (dy\ -\ C_i\ dx^i)^2\ ,\\
   \hat A^{(3)}_{ijk} &=& A^{(3)}_{ijk}\ ,\ \ \ \ \hat A^{(3)}_{yij}\ =\ 
\hat A^{(3)}_{jyi}\ =\ \hat A^{(3)}_{ijy}\ =\ B_{ij}\ .\ea
(\ref{wed}) is the (bosonic part of the) supermembrane action of \cite{tow}. 
It was not clear a priori that the effective action
for the type IIA membrane soliton is the standard eleven-dimensional 
supermembrane action,\footnote{E.g., an 
optimist might have hoped to find that M-theory is a topological field theory.}
but here we have derived it directly from D-branes.

Likewise, it should be possible to directly derive from D-branes 
the effective action of
M-branes that themselves have Dirichlet boundaries.
Anomaly cancellation arguments suggest that if the boundary surface lies on 
a 9-brane, an $E_8$ gauge theory appears on the 9-brane
\cite{horava}. There does not seem to be an obvious candidate for 
this $E_8$ gauge theory.
However, it seems that a Kalb-Ramond field must live on the 9-brane:
M-theory has a gauge field $A^{(3)}_{\mu\nu\sigma}$ with gauge invariance
$$A^{(3)}_{\mu\nu\sigma}\ \rightarrow\ A^{(3)}_{\mu\nu\sigma}\ 
+\ \partial_{[\mu}\Lambda_{\nu\sigma]}\ .$$
If the M-brane $M$ has a Dirichlet boundary, 
then gauge invariance is violated by a surface term:
$$\delta\int_Md^3x\ A^{(3)}_{\mu\nu\sigma}\ \epsilon^{\alpha\beta\gamma}\
  \partial_\alpha x^\mu \partial_\beta x^\nu \partial_\gamma x^\sigma\
  =\ \oint_{\partial M} \Lambda_{\mu\nu}\ \epsilon^{\alpha\beta}\ 
\partial_\alpha x^\mu\partial_\beta x^\nu\
.$$
To recover gauge invariance, a gauge field $A_{\mu\nu}$ 
with transformation law $\delta A = -\delta\Lambda$,
$$ \oint_{\partial M} A_{\mu\nu}\ \epsilon^{\alpha\beta}\ 
\partial_\alpha x^\mu\partial_\beta x^\nu\ ,$$
must be introduced on the boundary surface.
Of course, along with $A_{\mu\nu}$ one could introduce a complete set of 
string fields on the boundary surface of the M-brane.
Clarification of these issues is left for future work.

 {}\section*{5. Conclusions}

To summarize, effective actions for superstring solitons 
can be derived from Dirichlet branes.
Our results were obtained after integrating out (or rather, dualizing)
the boundary 
gauge field that appears in the formalism.
In the case of the one-branes of the type IIB theory, this
can be used to explicitly construct dual perturbation expansions 
of string theory with
elementary strings replaced by solitons, thus further confirming strong-weak 
coupling duality.
In the case of the two-branes of the type IIA theory, one finds that 
there is a hidden eleventh target space coordinate and that the two-brane 
action, which is also the M-brane action,
is the Bergshoeff-Sezgin-Townsend supermembrane action. 
Considering Dirichlet two-branes with boundaries might
contribute to a better understanding of M-theory.

The effective world-brane actions are easily extended to 
higher branes -- the 3-,5-,7- and 9-branes
of the type IIB theory and the 4-,6- and 8-branes of the type IIA theory. 
Of course one does not know how to
make sense of these branes quantum mechanically;
if one introduces, say, a four-dimensional dynamical metric as a step
towards quantizing three-branes,
one meets the old problem that four-dimensional gravity with only second-order
derivative terms in the action is not renormalizable (at least perturbatively). 
Fourth-order derivative gravity can be
renormalizable but not unitary -- there are negative norm states on 
the three-brane,
if its signature is chosen to be Minkowkian.
Perhaps further study of D-branes can suggest a way out.

\vskip 10mm
{\large\bf Acknowledgements}

I would like to thank C. Callan, D. Gross, P. Horava, I. Klebanov, 
J. Polchinski, A. Polyakov and A. Tseytlin for conversations.
I have also benefited from E. Witten's lectures on duality in Princeton.
This work is supported in part by NSF 
grant PHY90-21984 and by a grant from the Packard foundation.

{}

\end{document}